\documentclass[12pt]{article} 
\font\smallit=cmti10 
\font\smalltt=cmtt10 
\font\smallrm=cmr9

\usepackage[pdftex,breaklinks,colorlinks]{hyperref}
\hypersetup{  % hyperref configuration, metadata
  pdftitle =
    {Efficient model chemistries for peptides. II. Basis set convergence in the B3LYP method.},
  pdfauthor =
    {Pablo Echenique},
  pdfsubject=
    protein folding,
  pdfkeywords =
    protein folding ab initio quantum chemistry B3LYP dipeptide basis set,
}
\hypersetup{  % hyperref configuration, colors
    citecolor = blue,
    urlcolor = blue
}

\usepackage{amstext,amsmath,amssymb} % AMS Math
\usepackage{txfonts}                 % Public Times New Roman text & math font

\usepackage{graphicx}
\usepackage{epsfig}
\usepackage[]{color}

\usepackage[square,comma,numbers,sort&compress]{natbib}
\usepackage{hypernat}  % For natbib to interact with hyperref

\begin{document} 
\vspace*{-40pt} 
\begin{center} {\bf{Efficient model chemistries for peptides. II.\\
 Basis set convergence in the B3LYP method.}}
\vskip 7pt 
{\bf Pablo ECHENIQUE}\\
{\smallit Instituto de Biocomputaci{\'o}n y F{\'{\i}}sica de
Sistemas Complejos (BIFI),\\
and Departamento de F{\'{\i}}sica Te{\'o}rica, Universidad de Zaragoza, \\
Pedro Cerbuna 12, E-50009 Zaragoza, Spain}\\
E-mail: {\tt echenique.p@gmail.com}
\vskip 7pt 
{\bf Gregory A. CHASS}\\
{\smallit Global Institute Of COmputational Molecular and Materials Science
(GIOCOMMS),\\ 
and School of Chemistry, University of Wales, Bangor, Gwynedd, LL57
2UW United Kingdom, \\
and College of Chemistry, Beijing Normal University, Beijing, 100875, China}
\end{center} 
\vskip 5pt 
\leftline{\smallrm PACS: 07.05.Tp; 31.15.Ar; 31.50.Bc; 87.14.Ee; 87.15.Aa; 89.75.-k}
\vskip -5pt 
{\smallrm Keywords: peptides, quantum chemistry, PES, B3LYP, basis set convergence}
\vskip 10pt 

\centerline{\bf Abstract}
{\footnotesize 
\noindent
  Small peptides are model molecules for the amino acid residues that are the
  constituents of proteins. In any bottom-up approach to understand the
  properties of these macromolecules essential in the functioning of every
  living being, to correctly describe the conformational behaviour of small
  peptides constitutes an unavoidable first step. In this work, we present an
  study of several potential energy surfaces (PESs) of the model dipeptide
  HCO-{\small L}-Ala-NH$_2$. The PESs are calculated using the B3LYP
  density-functional theory (DFT) method, with Dunning's basis sets cc-pVDZ,
  aug-cc-pVDZ, cc-pVTZ, aug-cc-pVTZ, and cc-pVQZ. These calculations, whose
  cost amounts to approximately 10 years of computer time, allow us to study
  the basis set convergence of the B3LYP method for this model peptide.  Also,
  we compare the B3LYP PESs to a previous computation at the
  MP2/6-311++G(2df,2pd) level, in order to assess their accuracy with respect
  to a higher level reference. All data sets have been analyzed according to a
  general framework which can be extended to other complex problems and which
  captures the nearness concept in the space of model chemistries (MCs).}

\pagestyle{myheadings} 
\markright{\smalltt B3LYP basis set convergence in peptides : Echenique and Chass}
\thispagestyle{empty} 
\baselineskip=15pt 
%\vskip 30pt 

\section{Introduction} 
\label{sec:introduction}

In any bottom-up attempt to understand the behaviour of protein
molecules (in particular, the still elusive protein folding process
\cite{Ech2007COP,Anf1973SCI,Sko2005PNAS,Dag2003TBS,Hon1999JMB}), the
characterization of the conformational preferences of short peptides
\cite{Zho2006JCTC,Tor2006MP,DiS2005JCTC,Per2004JCC,Bea1997JACS,Heg2007PRL,Per2003CEJ,Els2000CP}
constitutes an unavoidable first step. Due to the lower numerical
effort required and also to the manageability of their conformational
space, the most frequently studied peptides are the shortest ones: the
\emph{dipeptides}
\cite{Csa1999PBMB,Lan2005PSFB,Koo2002JPCA,Hud2001JCC}, in which a
single amino acid residue is capped at both the N- and C-termini with
neutral peptide groups. Among them, the most popular choice has been
the \emph{alanine} dipeptide
\cite{Ech2008JCC,Ros1979JACS,Mez1985JACS,Hea1989IJQC,Per1991JACS,Hea1991JACS,Fre1992JACS,Gou1994JACS,Bea1997JACS,End1997JMST,Rod1998JMST,Els2001CP,Yu2001JMS,Iwa2002JMST,Var2002JPCA,Per2003JCC,Wan2004JCC,Ech2006JCCb},
which, being the simplest chiral residue, shares many similarities
with most of the rest of dipeptides for the minimum computational
price.

Although classical force fields
\cite{Pon2003APC,Mac1998BOOK,Bro1983JCC,VGu1982MM,Cor1995JACS,Pea1995CPC,Jor1988JACS,Jor1996JACS,Hal1996JCCa}
are the only feasible choice for simulating large molecules at
present, they have been reported to yield inaccurate \emph{potential
energy surfaces} (PESs) for dipeptides
\cite{Mac2004JCC,Mac2004JACS,Kan2002JMST,Kam2001JPCB,Rod1998JMST} and
short peptides \cite{Wan2006JCTC,Bea1997JACS}. Therefore, it is not
surprising that they are widely recognized as being unable to
correctly describe the intricacies of the whole protein folding
process
\cite{Sno2005ARBBS,Sch2005SCI,Gin2005NAR,Mac2004JCC,Mor2004PNAS,Gom2003BOOK,Kar2002NSB,Bon2001ARBBS}.
On the other hand, albeit prohibitively demanding in terms of
computational resources, ab initio quantum mechanical calculations
\cite{Cra2002BOOK,Jen1998BOOK,Sza1996BOOK} are not only regarded as
the correct physical description that in the long run will be the
preferred choice to directly tackle proteins (given the exponential
growth of computer power and the advances in the search for pleasantly
scaling algorithms \cite{Ech2007MP,Sha2006PCCP}), but they are also used in
small peptides as the reference against which less accurate methods
must be compared
\cite{Mau2007JCTC,Arn2006JPCB,Mac2004JCC,Mac2004JACS,Kam2001JPCB,Rod1998JMST,Bea1997JACS}
in order to, for example, parameterize improved generations of
additive, classical force fields for polypeptides.

However, despite the sound theoretical basis, in practical quantum
chemistry calculations a plethora of approximations must be typically
made if one wants to obtain the final results in a reasonable human
time. The exact `recipe' that includes all the assumptions and steps
needed to calculate the relevant observables for any molecular system
has been termed \emph{model chemistry} (MC) by John Pople. In his own
words, a MC is an ``approximate but well-defined general and
continuous mathematical procedure of simulation''
\cite{Pop1999RMP}.

After assuming that the particles involved move at non-relativistic velocities
and that the greater weight of the nuclei allows to perform the
Born-Oppenheimer approximation, we are left with the problem of solving the
non-relativistic electronic Schr\"odinger equation \cite{Ech2007MP}.  The two
starting approximations to its exact solution that a MC must contain are,
first, the truncation of the $N$-electron space (in wave\-func\-tion-based
methods) or the choice of the functional (in DFT) and, second, the truncation
of the one-electron space, via the LCAO scheme (in both cases). The extent up
to which the first truncation is carried (or the functional chosen in the case
of DFT) is commonly called the \emph{method} and it is denoted by acronyms
such as RHF, MP2, B3LYP, CCSD(T), FCI, etc., whereas the second truncation is
embodied in the definition of a finite set of atom-centered Gaussian functions
termed \emph{basis set}
\cite{Ech2007MP,Gar2003BOOK,Jen1998BOOK,Sza1996BOOK,Hel1995BOOK}, which is
also designated by conventional short names, such as 6-31+G(d), TZP or
cc-pVTZ(--f). If we denote the method by a capital $M$ and the basis set by a
$B$, the specification of both is conventionally denoted by $L:=M/B$ and
called a \emph{level of the theory}. Typical examples of this are RHF/3-21G or
MP2/cc-pVDZ \cite{Cra2002BOOK,Jen1998BOOK,Sza1996BOOK}.

Note that, apart from these approximations, which are the most commonly used
and the only ones that are considered in this work, the MC concept may include
a lot of additional features: the heterolevel approximation (explored in a
previous work in this series \cite{Ech2008JCC}), protocols for extrapolating
to the infinite-basis set limit
\cite{Jur2006PCCP,Pet2005JCP,Jen2005TCA,Li2001CPL,Nyd1981JCP}, additivity
assumptions \cite{Jur2002CPL,Ign1991JCC,Dew1989JCC,Nob1982CPL}, extrapolations
of the M{\o}ller-Plesset series to infinite order \cite{Pop1983IJQC}, removal
of the so-called \emph{basis set superposition error} (BSSE)
\cite{Cre2005JCP,Sen2001IJQC,May1998JCP,Jen1996CPL,May1987TCA,Boy1970MP,Jan1969CPL},
etc. The reason behind most of these techniques being the urging need to
reduce the computational cost of the calculations.

Now, although general applicability is a requirement that all MCs must
satisfy, general accuracy is not mandatory. Actually, the fact is that the
different procedures that conform a given MC are typically parameterized and
tested in very particular systems, which are often small molecules. Therefore,
the validity of the approximations outside that native range of problems must
be always questioned and checked.  However, while the approximate
computational cost of a given MC for a particular system is rather easy to
predict on the basis of simple scaling relations, its expected accuracy on a
particular problem could be difficult to predict a priori, specially if we are
dealing with large molecules in which interactions in very different energy
scales are playing a role. The description of the conformational behaviour of
peptides (or, more generally, flexible organic species), via their PESs in
terms of the soft internal coordinates, is one of such problems and the one
that is treated in this work.

To this end, we first describe, in sec.~\ref{sec:methods}, the
computational and theoretical methods used throughout the rest of the
document. Then, in sec.~\ref{sec:general_framework}, we introduce a
basic framework that rationalizes the actual process of evaluating the
efficiency of any MC for a complex problem. These general ideas are
used, in sec.~\ref{sec:results}, to perform an study of the
\emph{density-functional theory} (DFT) B3LYP
\cite{Ste1994JPC,Bec1993JCP,Lee1988PRB,Vos1980CJP} method with the
cc-pVDZ, aug-cc-pVDZ, cc-pVTZ, aug-cc-pVTZ, and cc-pVQZ Dunning's
basis sets \cite{Dun1989JCP,Ken1992JCP}.  To this end, we apply these
levels of the theory to the calculation the PES of the model dipeptide
HCO-{\small L}-Ala-NH$_2$ (see fig.~\ref{fig:num_ala_rama}), and
assess their efficiency by comparison with a reference PES. Finally,
in sec.~\ref{sec:conclusions}, the most important conclusions are
briefly summarized.

\section{Methods} 
\label{sec:methods}

All ab initio quantum mechanical calculations have been performed using the
GAMESS-US program \cite{Gor2005BOOK,Sch1993JCC} under Linux and on 2.2 GHz
PowerPC 970FX machines with 2 GB RAM memory.

\begin{figure}[t!]
\begin{center}
\includegraphics[scale=0.10]{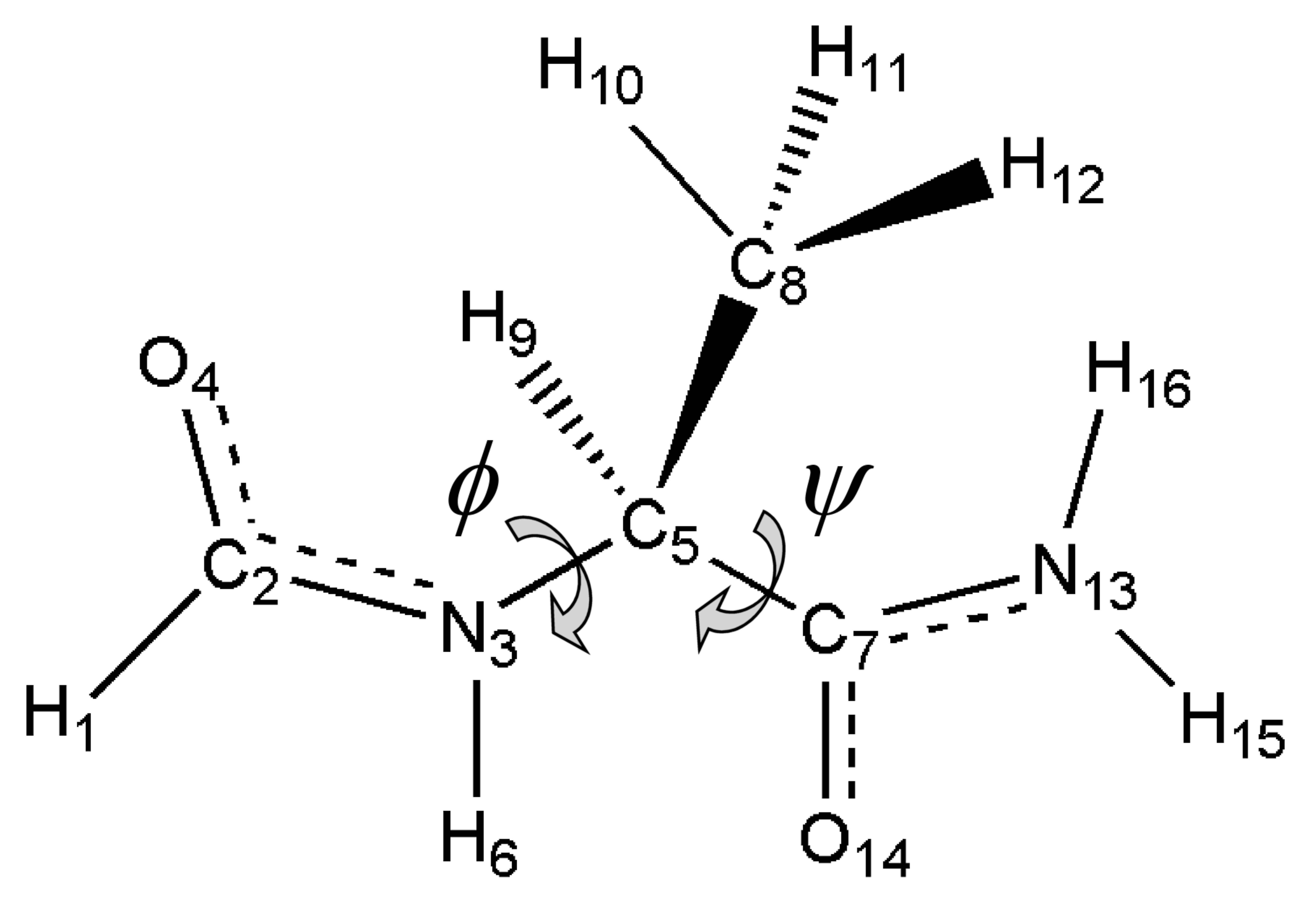}
\end{center}
\caption{\label{fig:num_ala_rama} {\small Atom numeration of the protected
dipeptide HCO-{\small L}-Ala-NH$_2$ according to the SASMIC scheme
introduced in ref.~\citealp{Ech2006JCCa}. The soft Ramachandran
angles $\phi$ and~$\psi$ are also indicated.}}
\end{figure}

The internal coordinates used for the Z-matrix of the HCO-{\small
  L}-Ala-NH$_2$ dipeptide in the GAMESS-US input files are the
\emph{Systematic Approximately Separable Modular Internal Coordinates}
(SASMIC) ones introduced in ref.~\citealp{Ech2006JCCa}. They are presented in
table~\ref{tab:coor_ala_rama} (see also fig.~\ref{fig:num_ala_rama} for
reference).

\begin{table}[!t]
\begin{center}
\begin{tabular}{cccc}
Atom name & Bond length & Bond angle & Dihedral angle \\
\hline\\[-8pt]
H$_{1}$ & & & \\
C$_{2}$ & (2,1) & & \\
N$_{3}$ & (3,2) & (3,2,1) & \\
O$_{4}$ & (4,2) & (4,2,1) & (4,2,1,3) \\
C$_{5}$ & (5,3) & (5,3,2) & (5,3,2,1) \\
H$_{6}$ & (6,3) & (6,3,2) & (6,3,2,5) \\
C$_{7}$ & (7,5) & (7,5,3) & $\phi:=${\bf (7,5,3,2)} \\
C$_{8}$ & (8,5) & (8,5,3) & (8,5,3,7) \\
H$_{9}$ & (9,5) & (9,5,3) & (9,5,3,7) \\
H$_{10}$ & (10,8) & (10,8,5) & (10,8,5,3) \\
H$_{11}$ & (11,8) & (11,8,5) & (11,8,5,10) \\
H$_{12}$ & (12,8) & (12,8,5) & (12,8,5,10) \\
N$_{13}$ & (13,7) & (13,7,5) & $\psi:=${\bf (13,7,5,3)} \\
O$_{14}$ & (14,7) & (14,7,5) & (14,7,5,13) \\
H$_{15}$ & (15,13) & (15,13,7) & (15,13,7,5) \\
H$_{16}$ & (16,13) & (16,13,7) & (16,13,7,15)
\end{tabular}
\end{center}
\caption{\label{tab:coor_ala_rama}{\small Internal coordinates in
Z-matrix form of the protected dipeptide HCO-{\small L}-Ala-NH$_2$
according to the SASMIC scheme introduced in
ref.~\citealp{Ech2006JCCa}. The numbering of the atoms is that
in fig.~\ref{fig:num_ala_rama}, and the soft Ramachandran angles
$\phi$ and~$\psi$ are indicated.}}
\end{table}

All PESs in this study have been discretized into a regular 12$\times$12 grid
in the bidimensional space spanned by the Ramachandran angles $\phi$ and
$\psi$, with both of them ranging from $-165^{\mathrm{o}}$ to
$165^{\mathrm{o}}$ in steps of $30^{\mathrm{o}}$. To calculate the PES at a
particular level of the theory, we have run constrained energy optimizations
at each point of the grid, freezing the two Ramachandran angles $\phi$ and
$\psi$ at the corresponding values. In order to save computational resources,
the starting structures were taken, when possible, from PESs previously
optimized at a lower level of the theory.  All the basis sets used in the
study have been taken from the GAMESS-US internally stored library, and
spherical Gaussian-type orbitals (GTOs) have been preferred, thus having 5
d-type and 7 f-type functions per shell.

We have computed 5 PESs, using the DFT B3LYP
\cite{Ste1994JPC,Bec1993JCP,Lee1988PRB,Vos1980CJP} method with the
cc-pVDZ, aug-cc-pVDZ, cc-pVTZ, aug-cc-pVTZ, and cc-pVQZ Dunning's
basis sets \cite{Dun1989JCP,Ken1992JCP}. The total cost of these
calculations in the machines used is around 10 years of computer time.

Also, let us note that the correcting terms to the PES coming from mass-metric
tensors determinants and from the determinant of the Hessian matrix have been
recently shown to be relevant for the conformational behaviour of peptides
\cite{Ech2006JCCb}. (The latter may be regarded as a residual entropy arising
from the elimination of the hard coordinates from the description.) Although,
in this study, we have included none of these terms, the PES calculated here
is the greatest part of the effective free energy \cite{Ech2006JCCb}, so that
it may be considered as the first ingredient for a further refinement of the
study in which the correcting terms are taken into account. The same may be
said about another important source of error in the calculation of relatives
energies in peptide systems: the already mentioned BSSE \cite{Var2002JPCA}.

In order to compare the PESs produced by the different MCs, a statistical
criterium (distance) introduced in ref.~\citealp{Alo2006JCC} has been used. Let
us recall here that this \emph{distance}, denoted by $d_{12}$, profits from
the complex nature of the problem studied to compare any two different
potential energy functions, $V_{1}$ and $V_{2}$. From a working set of
conformations (in this case, the 144 points of each PES), it statistically
measures the typical error that one makes in the \emph{energy differences} if
$V_{2}$ is used instead of the more accurate $V_{1}$, admitting a linear
rescaling and a shift in the energy reference.

Despite having energy units, the quantity $d_{12}$ approximately
presents all properties characteristic of a typical mathematical
metric in the space of MCs (hence the word `distance'), such as the
possibility of defining a symmetric version of it and a fulfillment of
the triangle inequality (see ref.~\citealp{Alo2006JCC} for the technical
details and sec.~\ref{sec:general_framework} for more about the
importance of these facts). It also presents better properties than
other quantities customarily used to perform these comparisons, such
as the energy RMSD, the average energy error, etc., and it may be
related to the Pearson's correlation coefficient $r_{12}$ by
\begin{equation}
\label{eq:d}
d_{12} = \sqrt{2}\,{\sigma}_{2}(1-r_{12}^{2})^{1/2} \  ,
\end{equation}
where $\sigma_2$ is the standard deviation of $V_2$ in the
working set.

Moreover, due to its physical meaning, it has been argued in
ref.~\citealp{Alo2006JCC} that, if the distance between two
different approximations of the energy of the same system is less than
$RT$, one may safely substitute one by the other without altering the
relevant dynamical or thermodynamical behaviour. Consequently, we
shall present the results in units of $RT$ (at
\mbox{$300^{\mathrm{o}}$ K}, so that $RT\simeq 0.6$ kcal/mol).

Finally, if one assumes that the effective energies compared will be
used to construct a polypeptide potential and that it will be designed
as simply the sum of mono-residue ones (more complex situations may be
found in real problems \cite{Ech2008UNPb}), then, the number
$N_{\mathrm{res}}$ of residues up to which one may go keeping the
distance $d_{12}$ between the two approximations of the the
$N$-residue potential below $RT$ is \cite{Alo2006JCC}
\begin{equation}
\label{eq:chPES_Nres}
N_{\mathrm{res}}=\left ( \frac{RT}{d_{12}} \right )^{2} \ .
\end{equation}

According to the value taken by $N_{\mathrm{res}}$ for a
comparison between a fixed reference PES, denoted by $V_1$, and a
candidate approximation, denoted by $V_2$, we shall divide the whole
accuracy range in sec.~\ref{sec:results} in three regions
depending on the accuracy: the \emph{protein region}, corresponding to
\mbox{$0 < d_{12} \leq 0.1 RT$}, or, equivalently, to \mbox{$100 \leq
  N_{\mathrm{res}} < \infty$}; the \emph{peptide region},
corresponding to \mbox{$0.1 RT < d_{12} \leq RT$}, or \mbox{$1 \leq
  N_{\mathrm{res}} < 100$}; and, finally, the \emph{inaccurate
  region}, where \mbox{$d_{12} > RT$}, and even for a dipeptide it is
not advisable to use $V_2$ as an approximation to~$V_1$. Of course,
these are only approximate regions based on the general idea that
we are not interested on the dipeptides as a final system, but only
as a mean to approach protein behaviour from the botton-up. Therefore,
not only the error in the dipeptides must be measured, but it must
also be estimated how this discrepancy propagates to polypeptide
systems.

\section{General framework} 
\label{sec:general_framework}

The general abstract framework behind the investigation presented in this
study (and also implicitly behind most of the works found in the literature),
may be described as follows:

The objects of study are the \emph{model chemistries} defined by Pople
\cite{Pop1999RMP} and discussed in the introduction. The MCs under scrutiny
are applied to a particular \emph{problem} of interest, which may be thought
to be formed by three ingredients: the \emph{physical system}, the
\emph{relevant observables} and the \emph{target accuracy}. The MCs are then
selected according to their ability to yield numerical values of the relevant
observables for the physical system studied within the target accuracy. The
concrete numerical values that one wants to approach are those given by the
\emph{exact model chemistry} MC$_\varepsilon$, which could be thought to be
either the experimental data or the exact solution of the non-relativistic
electronic Schr\"odinger equation \cite{Ech2007MP}.  However, the
computational effort needed to perform the calculations required by
MC$_\varepsilon$ is literally infinite, so that, in practice, one is forced to
work with a \emph{reference model chemistry} MC$^\mathrm{ref}$, which, albeit
different from MC$_\varepsilon$, is thought to be close to it.  Finally, the
set of MCs that one wants to investigate are compared to MC$^\mathrm{ref}$ and
the nearness to it is seen as approximating the nearness to MC$_\varepsilon$.

\begin{figure}[!b]
\begin{center}
\includegraphics[scale=0.40]{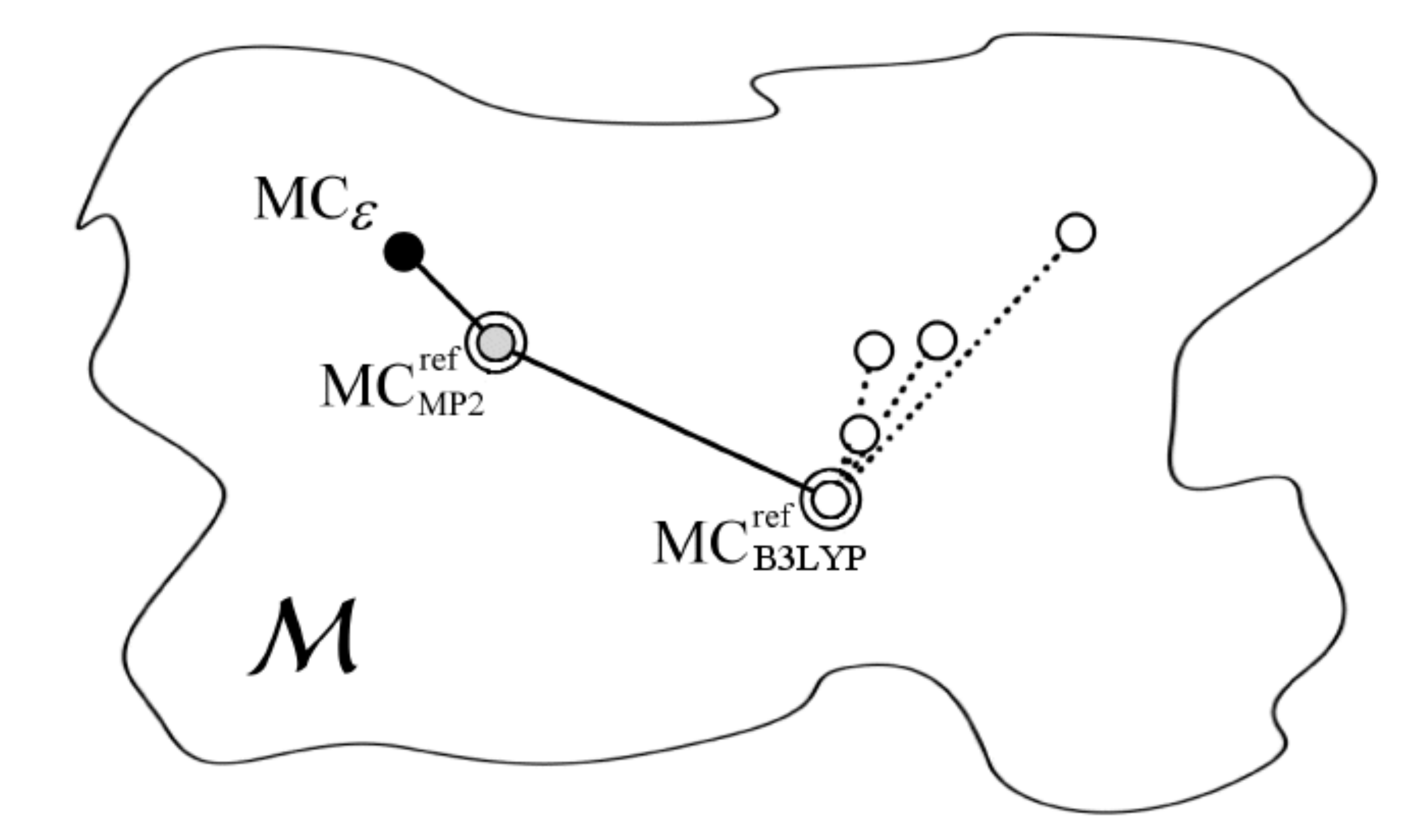}
\caption{\label{fig:mc_space} {\small Space $\mathcal{M}$ of all model
  chemistries. The exact model chemistry MC$_\varepsilon$ is shown as
  a black circle, the MP2 reference MC is shown as a grey-filled
  circle, and B3LYP MCs as white-filled ones. Both reference PESs are
  indicated with an additional circle around the points. The situation
  depicted is (schematically) the one found in this study, assuming
  that MP2 is a more accurate method than B3LYP to account for the
  conformational preferences of peptide systems. The positions of the
  different MCs have no relevance, and only the relative measured
  distances among them are qualitatively depicted.}}
\end{center}
\end{figure}

These comparisons are commonly performed using a numerical quantity
$\mathcal{D}$ that is a function of the relevant observables. In order
for the intuitive ideas about relative proximity in the $\mathcal{M}$
space to be captured and the above reasoning to be meaningful, this
numerical quantity $\mathcal{D}$ must have some of the properties of a
mathematical distance. In particular, it is advisable that the
\emph{triangle inequality} is obeyed, so that, for any model chemistry
MC, one has that
\begin{subequations}
\label{eq:ti}
\begin{align}
& \mathcal{D}(\mathrm{MC}_\varepsilon,\mathrm{MC}) \leq
 \mathcal{D}(\mathrm{MC}_\varepsilon,\mathrm{MC}^\mathrm{ref}) +
\mathcal{D}(\mathrm{MC}^\mathrm{ref},\mathrm{MC}) \ ,
  \label{eq:ti_a} \\
\displaybreak[0]
&\mathcal{D}(\mathrm{MC}_\varepsilon,\mathrm{MC}) \geq
 \big| \mathcal{D}(\mathrm{MC}_\varepsilon,\mathrm{MC}^\mathrm{ref}) -
       \mathcal{D}(\mathrm{MC}^\mathrm{ref},\mathrm{MC}) \big| \ ,
  \label{eq:ti_b}
\end{align}
\end{subequations}
and, assuming that
$\mathcal{D}(\mathrm{MC}_\varepsilon,\mathrm{MC}^\mathrm{ref})$ is
small (and $\mathcal{D}$ is a positive function), we obtain
\begin{equation}
\label{eq:pess_ti3}
\mathcal{D}(\mathrm{MC}_\varepsilon,\mathrm{MC}) \simeq
  \mathcal{D}(\mathrm{MC}^\mathrm{ref},\mathrm{MC}) \ ,
\end{equation}
which is the sought result in agreement with the ideas stated at
the beginning of this section.

The distance $d_{12}$ introduced in ref.~\citealp{Alo2006JCC} and
summarized in the previous section, measured in this case on the
conformational energy surfaces (the relevant observable) of the model
dipeptide HCO-{\small L}-Ala-NH$_2$ (the physical system),
approximately fulfills the triangle inequality and thus captures the
\emph{nearness} concept in the space $\mathcal{M}$ of model
chemistries.

This space, $\mathcal{M}$, containing all possible MCs, is a rather complex
and multidimensional one. For example, two commonly used `dimensions' which
may be thought to parameterize $\mathcal{M}$ are the size of the basis set and
the amount of electron correlation in the model (or the quality of the DFT
functional used). However, since there are many ways in which the size of a
basis set or the electron correlation may be increased and there are
additional approximations that can be included in the MC definition (see
sec.~\ref{sec:introduction}), the `dimensions' of $\mathcal{M}$ can be
considered to be many more than two.

The definition of a distance, such as the one described in the
previous lines, for a given problem of interest helps to provide a
certain degree of structure into this complex space.
In fig.~\ref{fig:mc_space} a two-dimensional scheme of the overall
situation found in this study is presented.

\section{Results} 
\label{sec:results}

\begin{table}[b!]
\begin{center}
\begin{tabular}{l@{\hspace{20pt}}crrr}
 MCs & $d_{12}/RT$ $^{a}$ & $a_{12}$ $^{b}$ &
 \multicolumn{1}{c}{$N_{\mathrm{res}}$ $^{c}$} &
 $t$ $^{d}$ \\
\hline\\[-8pt]
 B3LYP/aug-cc-pVTZ & 0.079 &  15.2 & 159.8 & 79.09\%\\
 B3LYP/cc-pVTZ     & 0.191 &  21.1 &  27.4 &  9.78\%\\
 B3LYP/aug-cc-pVDZ & 0.172 &  82.8 &  33.7 &  5.27\%\\
 B3LYP/cc-pVDZ     & 1.045 & 109.4 &   0.9 &  1.29\%\\
\end{tabular}
\end{center}
\caption{\label{tab:bs_convergence}{\small Basis set convergence
    results for the B3LYP MCs investigated in this work.
    $^{a}$Distance with the B3LYP/cc-pVQZ reference in units of $RT$
    at \mbox{$300^{\mathrm{o}}$ K}. $^{b}$Energy offset with the
    reference MC in kcal/mol. $^{c}$Maximum number of residues
    in a polypeptide potential up to which the corresponding MC may
    correctly approximate the reference
    (under the assumptions in sec.~\ref{sec:methods}).
    $^{d}$Required computer time,
    expressed as a fraction of $t_{\mathrm{ref}}$.}}
\end{table}

Before starting with the results of the calculations, let us introduce the
concept of \emph{efficiency} of a particular MC that shall be used: It is
laxly defined as a balance between accuracy (in terms of the distance
introduced in sec.~\ref{sec:methods}) and computational cost (in terms of
computer time). It can be graphically extracted from the \emph{efficiency
  plots}, where the distance $d_{12}$ between any given MC and a reference one
is shown in units of $RT$ in the $x$-axis, while, in the $y$-axis, one can
find the computer time taken for each MC (see the following pages for two
examples). As a general thumb-rule, \emph{we shall consider a MC to be more
  efficient for approximating the reference when it is placed closer to the
  origin of coordinates in the efficiency plot}.  This approach is
intentionally non-rigorous due to the fact that many factors exist that
influence the computer time but may vary from one practical calculation to
another; such as the algorithms used, the actual details of the computers
(frequency of the processor, size of the RAM and cache memories, system bus
and disk access velocity, operating system, mathematical libraries, etc.), the
starting guesses for the SCF orbitals or the starting structures in geometry
optimizations.

Taking all this into account, the only conclusions that shall be drawn in this
work about the relative efficiency of the MCs studied are those deduced from
strong signals in the plots and, therefore, those that can be extrapolated to
future calculations; in other words, \emph{the small details shall be
  typically neglected}.

\begin{figure}
\begin{center}
\includegraphics[scale=0.42]{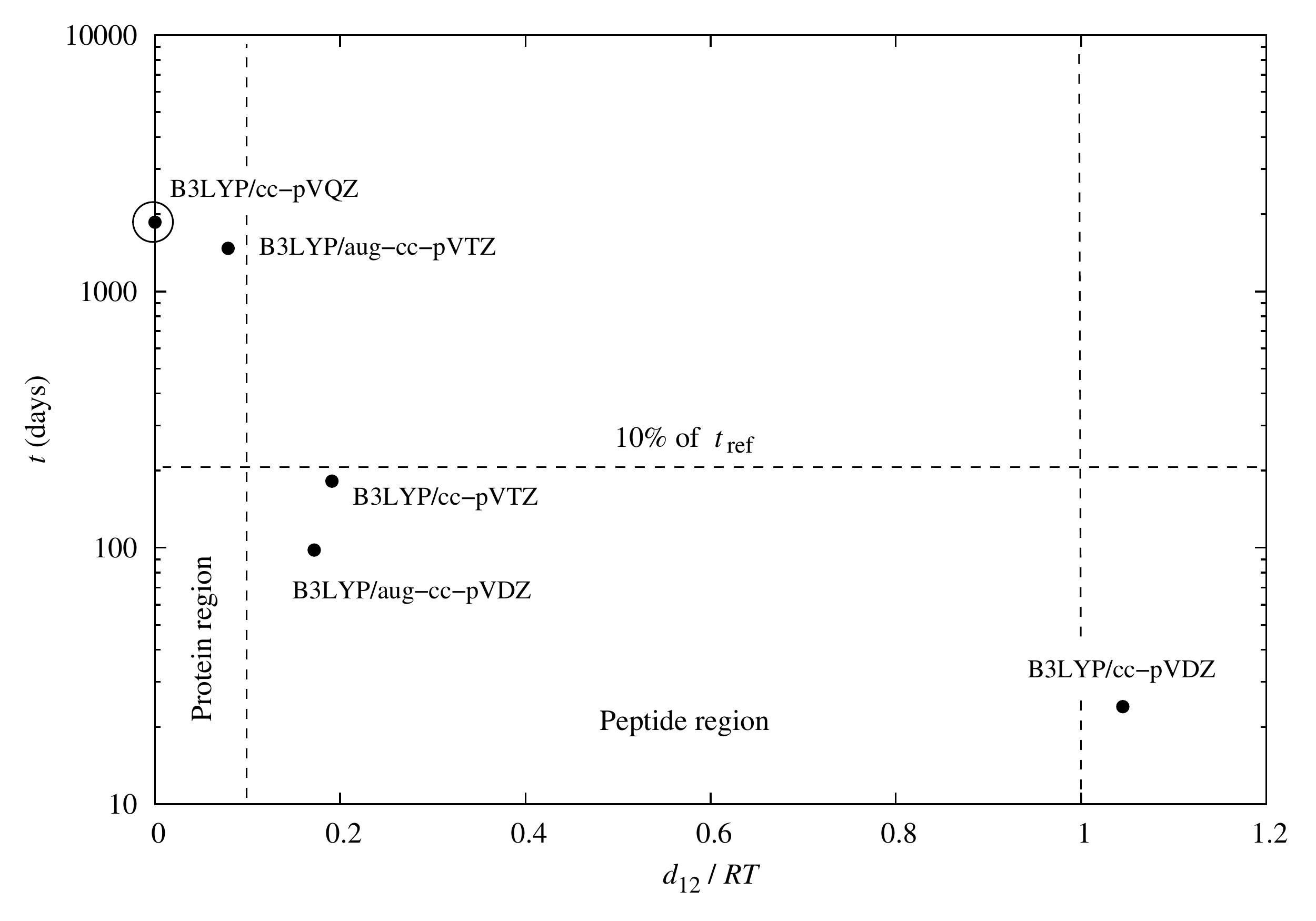}
\caption{\label{fig:efficiency_B3LYP}{\small Efficiency plot of all
  the B3LYP MCs studied. In the $x$-axis, we show the
  distance $d_{12}$, in units of $RT$ at \mbox{$300^{\mathrm{o}}$ K},
  between any given MC and the B3LYP/cc-pVQZ reference (indicated by an
  encircled point), while, in the $y$-axis, we present the
  computer time needed to compute the whole 12$\times$12 grid in the
  Ramachandran space of the model dipeptide HCO-{\small L}-Ala-NH$_2$.
  The different accuracy regions are labeled, and the 10\% of the
  time $t_{\mathrm{best}}$ taken by the reference MC is also
  indicated.}}
\end{center}
\end{figure}

\begin{figure}
\begin{center}
\includegraphics[scale=0.55]{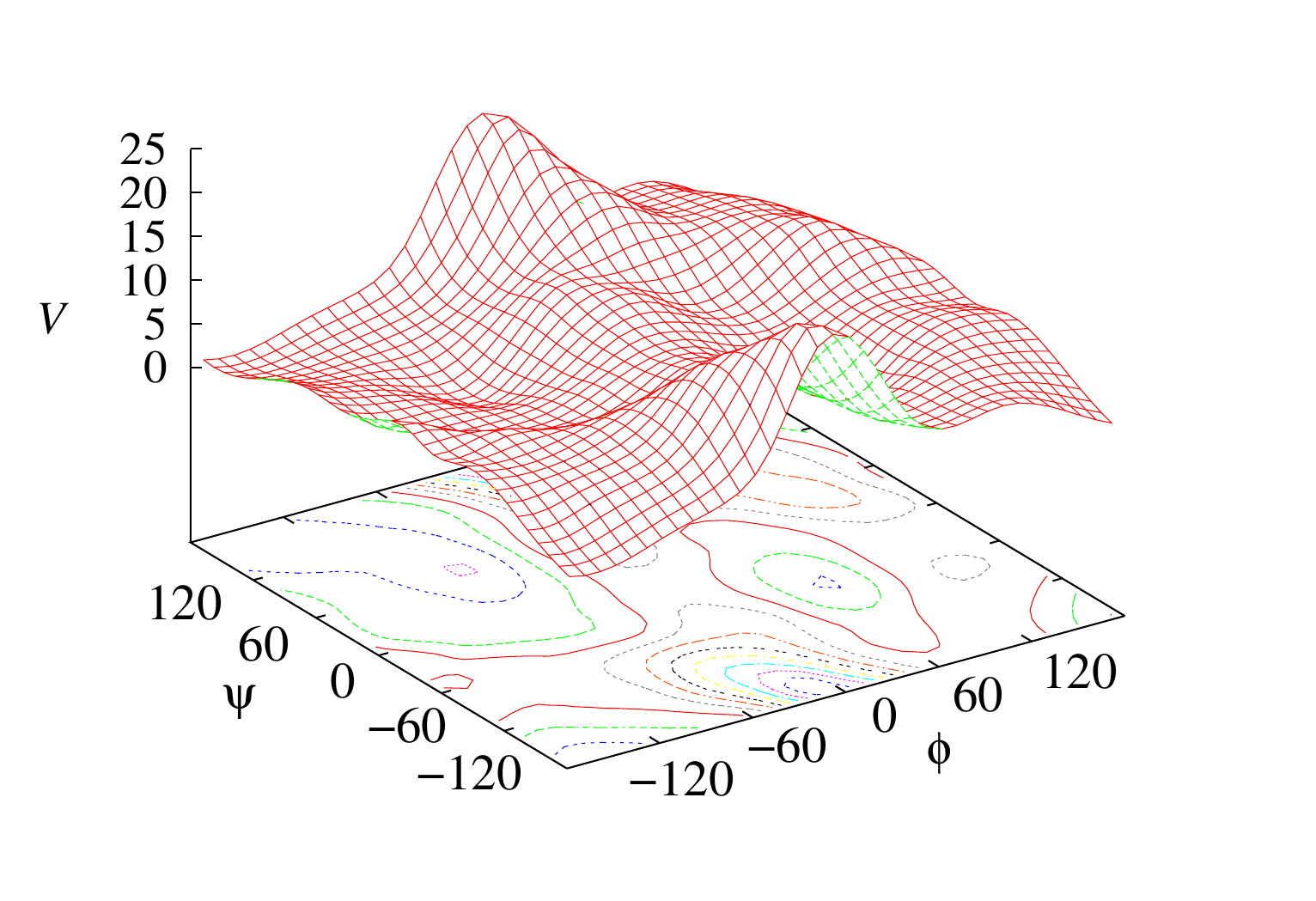}
\caption{\label{fig:best_PES_B3LYP} {\small Potential energy surface of the
model dipeptide HCO-{\small L}-Ala-NH$_2$ computed at the
B3LYP/cc-pVQZ level of the theory. The PES has been originally
calculated in a 12$\times$12 discrete grid in the space spanned by the
Ramachandran angles $\phi$ and $\psi$ and later smoothed with bicubic
splines for visual convenience. The energy reference has been set to
zero. (At this level of the theory, the absolute energy of the minimum
point in the 12$\times$12 grid, located at $(-75^o,75^o)$, is
$-417.199231353$ hartree).}}
\end{center}
\end{figure}

In the first part of the study, we compare all B3LYP MCs to the one
with the largest basis set, B3LYP/cc-pVQZ (the highest level of the
theory calculated for this work, depicted in
fig.~\ref{fig:best_PES_B3LYP}) using the distance introduced in
sec.~\ref{sec:methods}. All mentions to the accuracy of any given MC
in this part must be understood as relative to this
reference. However, it has been reported that MP2 is a superior method
to B3LYP to account for the conformational behaviour of peptide
systems \cite{Kam2007JCTC}. Therefore, the absolute accuracy of the
B3LYP MCs calculated here is probably closer to the relative accuracy
with respect to the MP2/6-311++G(2df,2pd) reference in what
follows. In this spirit, this part of the study should be regarded as
an investigation of the convergence to \emph{the infinite basis set
  B3LYP limit}, for which the best B3LYP MC here is probably a good
approximation.

The results are depicted in fig.~\ref{fig:efficiency_B3LYP}, and in
table~\ref{tab:bs_convergence}. We can extract several conclusions
from them:

\begin{itemize}
\item Regarding the convergence to the infinite basis set limit, we
  observe that only the most expensive MC, B3LYP/aug-cc-pVTZ,
  correctly approximates the reference for peptides of more than 100
  residues. On the other hand, for only 5.27\% of the computer time
  $t_{\mathrm{ref}}$ taken by the reference MC, we can use
  B3LYP/aug-cc-pVDZ, which correctly approximates it up to 30-residue
  peptides. Finally, the MC with the smallest basis set, B3LYP/cc-pVDZ
  cannot properly replace the reference even in dipeptides.

\item In ref.~\cite{Ech2008JCC}, using Pople's basis sets
  \cite{Dit1971JCP,Heh1972JCP,Har1973TCHIA,Fri1984JCP,Kri1980JCP,Bin1980JACS,Spi1982JCC,Cla1983JCC},
  we saw that ``the general rule that is sometimes assumed when
  performing quantum chemical calculations, which states that `the
  more expensive, the more accurate', is rather coarse-grained and
  relevant deviations from it may be found.'' We recognized that ``One
  may argue that this observation is due to the unsystematic way in
  which Pople basis sets can be enlarged and that the correlation
  between accuracy and cost will be much higher if, for example, only
  Dunning basis sets are used.'', which is definitely observed in
  fig.~\ref{fig:efficiency_B3LYP}, but we argued that this was
  something to be expected, since ``there are two few Dunning basis
  sets below a reasonable upper bound on the number of elements to see
  anything but a line in the efficiency plot''. In the results
  presented in this work, we can see that, even if the correlation
  between accuracy and cost is higher in the case of Dunning's basis
  sets than in the case of Pople's, due to the smaller number of the
  former, we can still observe that the thumb-rule `the more
  expensive, the more accurate' breaks also in this case, since the
  B3LYP/aug-cc-pVDZ MC is, at the same time, more accurate and less
  costly than B3LYP/cc-pVTZ. In general, this idea applies to all the
  approximations that a MC may contain (see the introduction for a
  partial list), and justifies the systematic search for the most
  efficient combination of them for a given problem. This work is our
  second step (ref.~\cite{Ech2008JCC} is the first one) in that path
  for the particular case of the conformational behaviour of peptide
  systems.

\item The observation in the previous point also suggests that
  it may be efficient to include diffuse functions (the `aug-' in
  aug-cc-pVDZ) in the basis set for this type of problems.

\item The error of the studied MCs regarding the differences
  of energy (as measured by $d_{12}$) is much smaller than the
  error in the absolute energies (as measured by $a_{12}$),
  suggesting that the largest part of the discrepancy must be
  a systematic one.
\end{itemize}

\begin{figure}
\begin{center}
\includegraphics[scale=0.42]{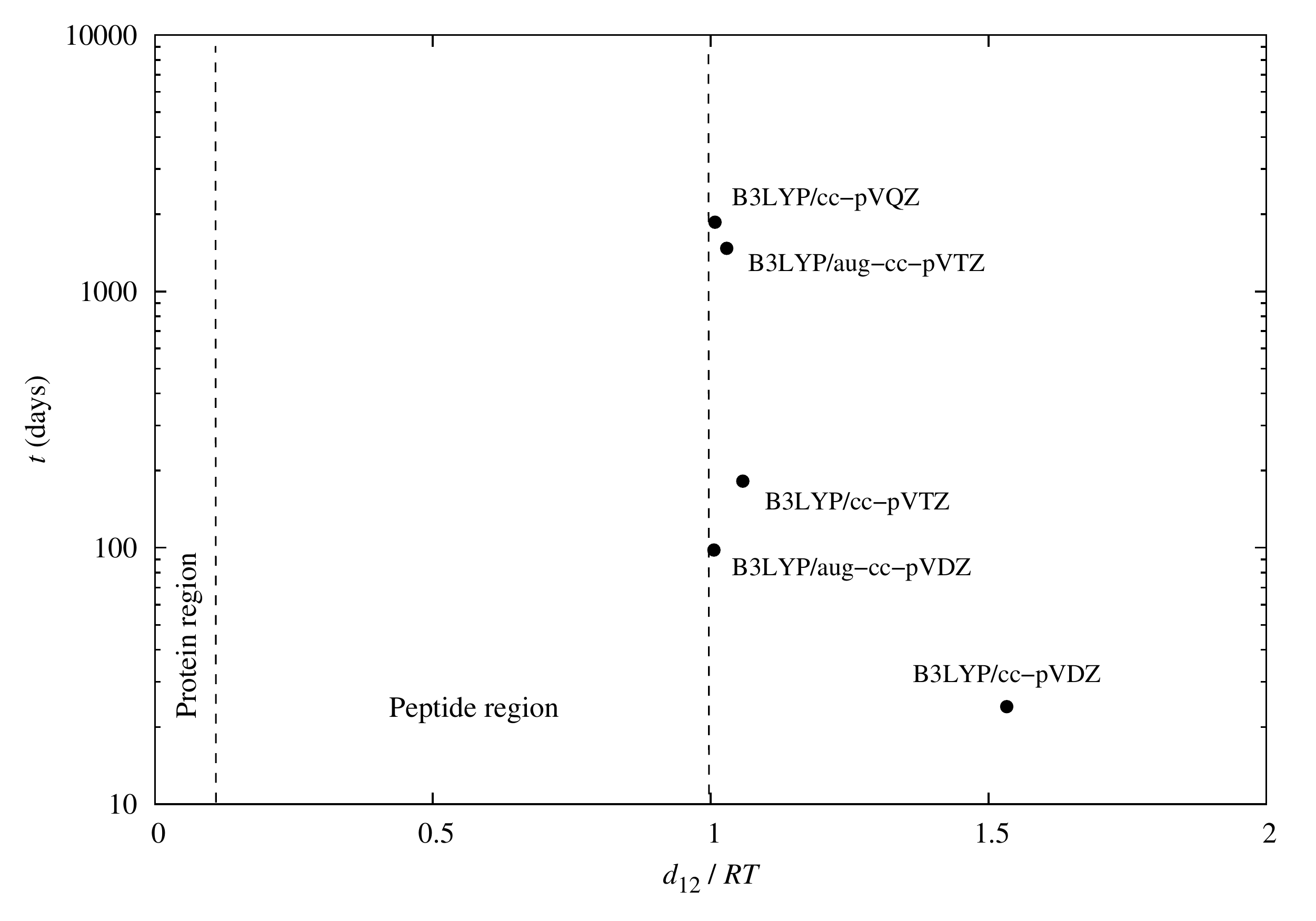}
\caption{\label{fig:efficiency_MP2}{\small Efficiency plot of all
  the B3LYP MCs studied. In the $x$-axis, we show the
  distance $d_{12}$, in units of $RT$ at \mbox{$300^{\mathrm{o}}$ K},
  between any given MC and the MP2/6-311++G(2df,2pd) reference calculated
  in ref.~\citealp{Ech2008JCC}, while, in the $y$-axis, we present the
  computer time needed to compute the whole 12$\times$12 grid in the
  Ramachandran space of the model dipeptide HCO-{\small L}-Ala-NH$_2$.
  The different accuracy regions are labeled}}
\end{center}
\end{figure}

In the second part of the study, we assess the absolute accuracy of
the B3LYP MCs by comparing them to the (as far as we are aware)
highest homolevel in the literature, the MP2/6-311++ G(2df,2pd) PES in
ref.~\cite{Ech2008JCC}. If one assumes that this level of the theory
may be close enough to the exact result for the given problem at hand,
then this comparison measures the `absolute' accuracy of the B3LYP
MCs, and not only their relative accuracy with respect to the B3LYP
infinite basis set limit, as we did in the previous part.  This is the
fundamental difference between figs.~\ref{fig:efficiency_B3LYP}
and~\ref{fig:efficiency_MP2}.

\begin{table}[!t]
\begin{center}
\begin{tabular}{l@{\hspace{20pt}}crrrr}
 MCs & $d_{12}/RT$ $^{a}$ & $a_{12}$ $^{b}$ &
 \multicolumn{1}{c}{$N_{\mathrm{res}}$ $^{c}$} & $t$ $^{d}$ \\
\hline\\[-8pt]
 B3LYP/cc-pVQZ     & 1.008 &  -457.2 & 0.98 & 1861 \\
 B3LYP/aug-cc-pVTZ & 1.029 &  -442.0 & 0.94 & 1472 \\
 B3LYP/cc-pVTZ     & 1.058 &  -436.1 & 0.89 &  182 \\
 B3LYP/aug-cc-pVDZ & 1.006 &  -374.4 & 0.99 &   98 \\
 B3LYP/cc-pVDZ     & 1.533 &  -347.8 & 0.43 &   24 \\
\end{tabular}
\end{center}
\caption{\label{tab:with_MP2}{\small Comparison of all the B3LYP MCs
    investigated in this work with the MP2/6-311++G(2df,2pd) in
    ref.~\citealp{Ech2008JCC}.
    $^{a}$Distance with the MP2/6-311++G(2df,2pd) reference in units of $RT$
    at \mbox{$300^{\mathrm{o}}$ K}. $^{b}$Energy offset with the
    reference MC in kcal/mol. $^{c}$Maximum number of residues
    in a polypeptide potential up to which the corresponding MC may
    correctly approximate the reference
    (under the assumptions in sec.~\ref{sec:methods}).
  $^{d}$Computer time needed for the calculation of the whole PES,
    in days.}}
\end{table}

The results of this part of the study are depicted in
fig.~\ref{fig:efficiency_MP2}, and in table~\ref{tab:with_MP2}. We can
extract several conclusions from them:

\begin{itemize}
\item All B3LYP MCs, including the largest one, B3LYP/cc-pVQZ, lie in
  the inaccurate region of the efficiency plot in
  fig.~\ref{fig:efficiency_MP2}, meaning that they cannot be reliably
  used to approximate the MP2/6-311++G(2df,2pd) reference even in the
  smallest dipeptides.
\item Related with the observations in the previous part of the
  study, we see that there is no point, if one is worried about
  absolute accuracy, in going beyond the aug-cc-pVDZ basis set in
  B3LYP.
\item The B3LYP/cc-pVDZ MC again performs significantly worse than
  the rest, agreeing with the results in the previous part of the
  study, and suggesting that cc-pVDZ may be a too small basis set
  for the problem tackled here.
\item Again, the error of the MCs in the differences
  of energy (as measured by $d_{12}$) is much smaller than the
  error in the absolute energies (as measured by $a_{12}$).
\end{itemize}

\section{Conclusions} 
\label{sec:conclusions}

In this study, we have investigated 5 PESs of the model dipeptide
HCO-{\small L}-Ala-NH$_2$, calculated with the B3LYP method, and the
cc-pVDZ, aug-cc-pVDZ, cc-pVTZ, aug-cc-pVTZ, and cc-pVQZ Dunning's
basis sets. We have first assessed the convergence of the B3LYP MCs to
the infinite basis set limit, and then we have evaluated their
absolute accuracy by comparing them to the (as far as we are aware)
highest homolevel in the literature, the MP2/6-311++G(2df,2pd) PES in
ref.~\cite{Ech2008JCC}. All the comparisons have been performed
according to a general framework which is extensible to further
studies, and using a distance between the different PESs that
correctly captures the nearness concept in the space of MCs. The
calculations performed here have taken around 10 years of computer
time.

The main conclusions of the study are the following:

\begin{itemize}
\item The complexity of the problem (the conformational behaviour of
  peptides) renders the correlation between accuracy and computational
  cost of the different quantum mechanical algorithms imperfect. This
  ultimately justifies the need for systematic studies, such as the
  one presented here, in which the most efficient MCs are sought for
  the particular problem of interest.
\item Assuming that the MP2/6-311++G(2df,2pd) level of the theory is
  closer to the exact solution of the non-relativistic electronic
  Schr\"odinger equation than B3LYP/cc-pVQZ, B3LYP is not a reliable
  method to study the conformational behaviour of peptides. Even if,
  as we emphasize at the end of this section, it may be dangerous to
  state that a method that performs well in the particular model of an
  alanine residue studied here will also be recommendable for longer
  and more complex peptides, we can clearly \emph{reject} any method
  that already fails in HCO-{\small L}-Ala-NH$_2$.
\item If B3LYP is still needed to be used, due to, for example,
  computational constraints, aug-cc-pVDZ represents a good compromise
  between accuracy and cost.
\item The error of the studied MCs regarding the differences
  of energy (as measured by $d_{12}$) is much smaller than the
  error in the absolute energies (as measured by $a_{12}$),
  suggesting that the largest part of the discrepancy must be
  a systematic one.
\end{itemize}

Finally, let us stress again that the investigation performed here have
used one of the simplest dipeptides. The fact that we have treated
it as an isolated system, the small size of its side chain and also
its aliphatic character, all play a role in the results obtained.
Hence, for bulkier residues included in polypeptides, and, specially
for those that contain aromatic groups, those that are charged or may
participate in hydrogen-bonds, the methods that have proved to be
efficient here must be re-tested and the conclusions drawn about the
B3LYP convergence to the infinite basis set limit, as well as those
regarding the comparison between B3LYP and MP2, should be
re-evaluated.

\section*{Acknowledgments}

\hspace{0.5cm} The numerical calculations in this work have been performed
thanks to a computer time grant at the Zaragoza node (Caesaraugusta) of the
Spanish Supercomputing Network (RES). We thank all the support staff there,
for the efficiency at solving the problems encountered. We also thank J. L.
Alonso for illuminating discussions.

This work has been supported by the research projects DGA (Arag\'on
Government, Spain) E24/3 and MEC (Spain)
\mbox{FIS2006-12781-C02-01}. P. Echenique is supported by a MEC (Spain)
postdoctoral contract.

{
\phantomsection
\addcontentsline{toc}{chapter}{References}
\bibliography{b3lyp}
}

\end{document}